\documentclass[letterpaper, 10 pt, conference]{ieeeconf}

\IEEEoverridecommandlockouts
\usepackage[english]{babel}

\usepackage{graphicx}
\usepackage[colorlinks=true, allcolors=blue]{hyperref}

\usepackage{amsmath,amssymb}       % blackboard math symbols
\usepackage{cases}
\usepackage{empheq}
\usepackage{mathbbol}
\usepackage{bm}
\usepackage{bbm}
\DeclareSymbolFontAlphabet{\amsmathbb}{AMSb}
\usepackage{mathtools}
\usepackage{accents}
\usepackage[colorinlistoftodos]{todonotes}

\usepackage{siunitx}
\usepackage{caption}
\usepackage{subcaption}
\usepackage{cite}

\usepackage{accents}
\usepackage{algorithm}
\usepackage{algpseudocode}

\graphicspath{./Figures/}

\newcommand{\SpaceofDistributions}{\mathcal{P}}
\newcommand{\StochasticKernel}{\ensuremath{T}}

\newcommand{\RVProcessNoise}{\ensuremath{W}}
\newcommand{\ProcessNoise}{\ensuremath{w}}
\newcommand{\SampleSet}{\Omega}
\newcommand{\SigmaAlg}{\mathcal{F}}
\newcommand{\ProbabilityMeasure}{\mathbb{P}}
\newcommand{\Sample}{\omega}

\newcommand{\SizePartition}{\ensuremath{N}}

\newcommand{\Powerset}[1]{2^{#1}}
\newcommand{\TargetSets}{\mathcal{T}}
\newcommand{\TargetSet}{\ensuremath{T}}
\newcommand{\SizeTargetSets}{M}
\newcommand{\TargetSetCollection}{\{\TargetSet_i\}_{i=1}^\SizeTargetSets}
\newcommand{\Partition}{\mathcal{Q}}
\newcommand{\SetInPartition}{\ensuremath{Q}}
\newcommand{\PartitionSetCollection}{\{\SetInPartition_i\}_{i=1}^\SizePartition}
\newcommand{\PartitionRelation}{\ensuremath{R}}
\newcommand{\MapToReference}{\ensuremath{\psi}}
\newcommand{\ReferencePointCluster}{\ensuremath{C}}
\newcommand{\ReferencePoint}{\ensuremath{c}}
\newcommand{\SubPartitionSize}{\ensuremath{L}}

\newcommand{\Flow}{\ensuremath{f}}
\newcommand{\RVStateTransSys}{\ensuremath{X}}
\newcommand{\StateTransSys}{\ensuremath{x}}
\newcommand{\DimState}{\ensuremath{n}}
\newcommand{\InputSetTransSys}{\ensuremath{U}}
\newcommand{\InputTransSys}{\ensuremath{u}}
\newcommand{\DimInput}{\ensuremath{m}}

\newcommand{\Domain}{\mathcal{X}}

\newcommand{\DiscreteStates}{\ensuremath{S}}
\newcommand{\DiscreteActions}{\ensuremath{\mathcal{A}}}
\newcommand{\StateAbs}{\ensuremath{s}}
\newcommand{\ActionAbs}{\ensuremath{a}}
\newcommand{\TransitionProbability}{P}
\newcommand{\IntervalTransitionProbability}{\TransitionProbability_\updownarrow}
\newcommand{\AggregatedTransitionProbability}{\TransitionProbability_{\fcolon}}

\newcommand{\NumSamples}{\ensuremath{Z}}
\newcommand{\MinProb}{\underline{p}}
\newcommand{\MaxProb}{\overline{p}}

\DeclareMathOperator{\Pre}{Pre}

\DeclareMathOperator*{\argmax}{arg\,max}
\DeclareMathOperator*{\argmin}{arg\,min}

\makeatletter
\newcommand{\fcolon}{%
  \mathrel{\mathpalette\fcolon@\relax}%
}
\newcommand{\fcolon@}[2]{%
  \sbox\z@{$\m@th#1:$}%
  \vbox to\ht\z@{%
    \hbox{$\m@th#1.$}%
    \vss
    \hbox{$\m@th#1.$}%
    \vss
    \hbox{$\m@th#1.$}%
  }%
}
\makeatother

\newcommand{\Reals}{\mathbb{R}}
\newcommand{\Naturals}{\mathbb{N}}

\newtheorem{definition}{Definition}
\newtheorem{theorem}{Theorem}
\newtheorem{assumption}{Assumption}
\newtheorem{example}{Example}

\newtheorem{remark}{Remark}

\begin{document}
%
% paper title
% Titles are generally capitalized except for words such as a, an, and, as,
% at, but, by, for, in, nor, of, on, or, the, to and up, which are usually
% not capitalized unless they are the first or last word of the title.
% Linebreaks \\ can be used within to get better formatting as desired.
% Do not put math or special symbols in the title.
\title{\LARGE \bf
Data-driven Interval MDP for Robust Control Synthesis
}
% tentative title
% alternatives: 
% interval aggregated MDP? 
% robust control synthesis ?

% author names and affiliations
% use a multiple column layout for up to three different
% affiliations

\author{Rudi Coppola$^{1}$, 
Andrea Peruffo$^{1}$, Licio Romao$^{2}$, Alessandro Abate$^{3}$ and Manuel Mazo Jr.$^{1}$
% \thanks{*This work was not supported by any organization}% <-this % stops a space
\thanks{$^{1}$R. Coppola, A. Peruffo and M. Mazo are with Faculty of Mechanical Engineering, 
TU Delft, Delft, The Netherlands. {\tt\small r.coppola@tudelft.nl}}%
\thanks{$^{2}$L. Romao is with the Dept of Aeronautics \& Astronautics, Stanford.}%
\thanks{$^{3}$A. Abate is with the Dept of Computer Science, University of Oxford.}
}

\maketitle
\thispagestyle{empty}
\pagestyle{empty}

\begin{abstract}
    The abstraction of dynamical systems is a powerful tool that enables the design of feedback controllers using a correct-by-design framework. We investigate a novel scheme to obtain data-driven abstractions of discrete-time stochastic processes in terms of richer discrete stochastic models, whose actions lead to nondeterministic transitions over the space of probability measures. The data-driven component of the proposed methodology lies in the fact that we only assume samples from an unknown probability distribution. We also rely on the model of the underlying dynamics to build our abstraction through backward reachability computations. The nondeterminism in the probability space is captured by a collection of Markov Processes, and we identify how this model can improve upon existing abstraction techniques in terms of satisfying temporal properties, such as safety or reach-avoid.
    The connection between the discrete and the underlying dynamics is made formal through the use of the scenario approach theory. Numerical experiments illustrate the advantages and main limitations of the proposed techniques with respect to existing approaches.
    
\end{abstract}

\IEEEpeerreviewmaketitle

%%%%%%%%%%%%%%%%%%%%%%%%%%%%%%%%%%%%%%%%%%%%%%%%%%%

\section{Introduction}
\label{sec:intro}

% -- abstractions of dynamical systems via TSs, MDPs 
The framework of control synthesis for dynamical systems usually includes a complex model, e.g. an ODE, coupled with a simple specification, e.g. stability, reachability, or invariance. 
These tasks are typically solved via a proxy approach as the construction of a Lyapunov function, or through numerical optimisation methods. 
Alternatively, one can abstract a dynamical system to a finite-state representation, typically in terms of an automaton or Markov decision process, for which much more complex specifications can be solved \cite{tab09,baier2008principles}.  
This process involves partitioning the state space into a finite set of regions, each represented by an abstract state, and computing transitions amongst abstract states, which is done by using a mathematical representation for the underlying dynamics. 
Actions in the abstract model correspond to control inputs (we refer the reader to \cite{tab09} for more details). Whenever the original dynamics is stochastic, the transitions of its discrete representation are probabilistic. 

% -- related works
A common modelling framework used in the stochastic context is provided by Markov decision processes (MDPs), which capture both the control synthesis task (i.e. policy synthesis) and the probabilistic nature of transitions. 
Richer frameworks, such as interval MDPs (iMDPs) \cite{givan2000bounded, dean1997model}, are employed to describe uncertain transition probabilities. 
%
% Hey Andrea, I am making some changes here. Feel free to ignore my comments if you find them useless. :) 
The evaluation of transition probabilities require prior knowledge on the stochastic nature of the underlying system, e.g., by evaluating the integral of the stochastic kernel over partitions, and may be computationally expensive to obtain. 
% -- citations on Girard Pappas / Ale / 
% -- scenario and  data-driven versions
% -- scenario PAC acronym 
As such, the use of samples for the construction of \emph{data-driven} abstractions has recently gained attention \cite{cubuktepe2020scenario, BRAPPSJ23, lavaei2022constructing,BRAJ23a,coppola2022data,coppola2023data,peruffo2022data,peruffo2023sampling, coppola2024datadriven, banse2023adaptive, banse2023data} both for deterministic and stochastic systems. 
These approaches consider mostly a black-box model and construct an abstraction from collected trajectories. Several of these works provide probably approximately correct (PAC) guarantees through the application of the scenario approach \cite{campi2011sampling, romao2022exact}. 
% Notably, the works \cite{thom} leverage the computation of the backward reachable set of the system dynamics and overlap it over the domain partitions. 

% -- contributions
\textbf{Contributions: } 
We revisit the approach presented in \cite{BRAJ23a} to abstract a discrete-time dynamical system with additive noise as an iMDP, using techniques from the scenario approach, with the overall goal of studying reach-avoid control problems. In doing so we introduce a Robust MDP where the ambiguity set has a particular structure. Building upon the results therein, we present a new strategy to construct such an abstraction by incorporating nondeterminism in the transitions: this allows us to search for policies over a larger action space and, therefore, to synthesise richer controllers for a wider variety of scenarios, with in particular the attainment of the specification of interest with a possibly higher probability, when compared to \cite{BRAJ23a}.

\textbf{Related Works: } 
Recently, a suite of techniques has emerged to build abstractions with data-driven techniques, mainly employing the scenario approach to provide PAC guarantees of correctness. 
In \cite{cubuktepe2020scenario, BRAPPSJ23, lavaei2022constructing}, Markov models are created using the scenario approach to evaluate transition probabilities in a stochastic dynamical model. In \cite{devonport2021symbolic}, a notion of PAC alternating simulation relationship is defined by sampling one-step state transitions.
%
% In \cite{makdesi2023data}, PAC over-approximations of monotone systems are employed to construct models; in \cite{kazemi2022data}, a sample-based growth rate is used to build an abstraction and synthesize a controller. 
%
Moving forward from one-step transitions are $\ell$-complete models (i.e. memory-based approaches), as in  \cite{coppola2022data,coppola2023data} for linear and nonlinear systems, and to synthesise controllers \cite{coppola2024datadriven}. 
These methods have been applied on event-triggered control models \cite{peruffo2022data,peruffo2023sampling}. 
Further, in \cite{banse2023data, banse2023adaptive} the construction of data-driven, memory-based models is equipped with an adaptive method to estimate the size of the needed memory from observations.

\section{Notation and Preliminaries}
\label{sec:notation}
%
%\licio{Of course, notation can be a very tricky subject. But here is the notations I tend to follow:
%\begin{itemize}
%    \item State space are denoted either as $\mathcal{X}$ or as $\Reals^\DimState$
%    \item The set of the probabilities measures over an abstract space is $\mathcal{P}(\mathcal{X})$. For continuous probability measures, I usually reserve the greek letters $\mu, \nu, \lambda$ -- the latter only if I want to refer to the Lebesgue measure. I also use the two boldface letters $\mathbb{P}$ and $\mathbb{Q}$. Sometimes, letters $p, q$ can also be used to refer to discrete probability measures.
%    \item To denote stochastic systems, as most of the time we are only reasoning abstractly with it, I prefer using the transition kernel notation. Using the $T: \mathcal{X} \rightarrow \mathcal{P}(\mathcal{X})$ as a measurable mapping between these relevant spaces.
%    \item I understand the presentation below is the one widely adopted to describe MDPs, and POMDPs, but I personally find it a bit misleading some times. I have recently written a paper with Thom that contains the notation I am suggesting here. I will share with you the paper and you can make up you mind about what notation to use in this paper. 
%\end{itemize}}

We denote by $\SpaceofDistributions(S)$ the set of all probability distributions on a discrete or continuous set $S$. Given a finite set $S$ we denote its power set as $\Powerset{S}$. We denote the interval defined by $a,b\in\Reals$ as  $[a,b]$.
A \emph{cover} of a set $\Domain$ is a finite collection of sets $\TargetSets=\TargetSetCollection$ such that each element of the collection is a subset of $\mathcal{X}$ and such that the union of the elements in the collection contains $\mathcal{X}$. A \emph{partition} of a set $\Domain$ is a cover $\Partition=\PartitionSetCollection$ such that the elements of the collection are pairwise disjoint.

%\textbf{ } \\ 
%\andrea{Conclude the various environments (definition, example, etc) with $\square$? }

%
\subsection{Stochastic Difference Equations} \label{sec:models}

Let us consider dynamical systems represented by a stochastic difference equation, where the dynamics of the state $\RVStateTransSys_{k+1} \in \Domain \subset \Reals^\DimState$ at time $k+1$ depends on a known function $\Flow: \Reals^\DimState \times \Reals^\DimInput \to \Reals^\DimState$ of the previous state and input, and on the noise $\RVProcessNoise_k$. We formally define the model below. 
\begin{definition}
    Consider a probability space $(\SampleSet,\SigmaAlg,\ProbabilityMeasure)$ and an independent and identically distributed random process $\{\RVProcessNoise_k(\Sample)\in\Reals^\DimState: k\in\Naturals_0, \Sample\in\SampleSet\}$.   A \emph{Stochastic Difference Equation} (SDE) with additive noise is a sequence of random variables defined as
    \begin{equation}\label{eq:add-noise}
        \RVStateTransSys_{k+1} = f(\RVStateTransSys_k,\InputTransSys_k) + \RVProcessNoise_k, 
    \end{equation}
    where $\InputTransSys_k:\Naturals_0\rightarrow\InputSetTransSys\subseteq\Reals^\DimInput$ and $\Flow:\Reals^\DimState\times\Reals^\DimInput\rightarrow\Reals^\DimState$. 
\end{definition}
 
% In view of \eqref{eq:add-noise} we 
Let us define a stochastic kernel $\StochasticKernel:\Reals^\DimState\times\InputSetTransSys\rightarrow\SpaceofDistributions(\Reals^\DimState)$ to describe the distribution of $\RVStateTransSys_{k+1}$ given $\StateTransSys_k$ and $\InputTransSys_k$ as
\begin{equation}\label{eq:stochastic-kernel}
    \RVStateTransSys_{k+1} \sim \StochasticKernel(\cdot \ | \ \StateTransSys_k, \InputTransSys_k).
\end{equation}

Furthermore we denote the next state under the \emph{nominal dynamics} of the SDE without additive noise as
\begin{equation}\label{eq:nominal-dyn}
    \hat{\StateTransSys}_{k+1} = f(\StateTransSys_k,\InputTransSys_k).
\end{equation}

Our goal is to produce an abstraction for $\eqref{eq:add-noise}$, where we assume to have full knowledge of the nominal dynamics $f(\cdot)$, whilst the  distribution of the noise $\RVProcessNoise_k$, and hence the distribution of $\RVStateTransSys_k$, is unknown.

%\begin{remark}
%    Assumption \ref{assump:non-degeneracy} is stronger than Assumption $4$ in \cite{romao2022exact}, which is required to show tightness of the sampling-and-discarding procedure used in the sequel. Details are omitted for brevity.
%\end{remark}

\subsection{Reach-avoid Specifications}
We focus on synthesising a controller for a SDE enforcing a reach-avoid specification over a finite time horizon. Let $\Domain_G\subset\Domain$ be a \emph{goal} set and let $\Domain_U\subset\Domain$ be an \emph{unsafe} set. A specification $\varphi_{x_0}^K$ is satisfied if the system, when initialized at $x_0$, reaches the goal set $\Domain_G$ within $K$ steps, while avoiding the unsafe set $\Domain_U$. 
% Due to the stochastic nature of a DTSDS, it is in general not possible to answer in absolute terms whether a specification is satisfied or not. 
Since the state of the system at time $k$, $\RVStateTransSys_k$, is a random variable we describe the \emph{probability} of satisfying a specification under a controller $\InputTransSys$, denoted $\ProbabilityMeasure^\InputTransSys\{\varphi^K_{x_0}\}$.

\smallskip
\textbf{Problem Statement}\emph{ Given a reach-avoid specification $\varphi^K_{x_0}$ and a SDE with \emph{unknown} additive noise, compute a controller $u$ and a lower bound on the probability of satisfying $\varphi^K_{x_0}$.}

\subsection{Markov Models}

This problem can be tackled by means of formal abstractions represented by Markov models. 
% namely, we first translate the DTSDS into a tailored Markov model, specifically an interval Markov decision process (iMDP); next, we find an abstract policy for the abstraction, and finally, we refine the policy for the abstraction to a controller for the DTSDS. 
In the following, we recall the notions related to Markov models that are required for our discussion, see \cite{givan2000bounded} for details.
\begin{definition}\label{def:mdp}[\cite{givan2000bounded}]
    A \emph{Markov Decision Process} (MDP) is a tuple $M = (\DiscreteStates,\DiscreteActions, \TransitionProbability, R)$ where $\DiscreteStates$ is a finite \emph{set of states}, $\DiscreteActions$ is a set of \emph{actions} where $\DiscreteActions(\StateAbs)$ indicates the enabled actions in $\StateAbs \in \DiscreteStates$, $\TransitionProbability:\DiscreteStates\times\DiscreteActions\rightarrow\SpaceofDistributions(\DiscreteStates)$ is a \emph{transition probability function} and $R:\DiscreteStates\rightarrow\mathbb{R}$ is a \emph{reward function}.
\end{definition}
\smallskip 
\begin{definition}[\cite{givan2000bounded}]
    An \emph{Interval Markov Decision Process} (iMDP) is a tuple $M_\updownarrow=(\DiscreteStates,\DiscreteActions,\IntervalTransitionProbability,R)$ where $\DiscreteStates$ and $\DiscreteActions$ are defined as in Definition \ref{def:mdp},  $\IntervalTransitionProbability:\DiscreteStates\times\DiscreteActions\rightrightarrows \SpaceofDistributions(\DiscreteStates)$ is an \emph{uncertain transition probability function} such that for all $\StateAbs$, $\StateAbs'$ and $\ActionAbs$ there exists $0\leq a\leq b\leq 1$ such that $\IntervalTransitionProbability(\StateAbs,\ActionAbs)(\StateAbs')= [a,b]$, and $R:\DiscreteStates\rightarrow\Reals$ is a reward function. 
\end{definition}

\smallskip

\begin{definition}%[Aggregated Markov Decision Process]
    A \emph{Robust Markov Decision Process} (RMDP) is a tuple $M_{\fcolon}=(\DiscreteStates,\DiscreteActions,\AggregatedTransitionProbability,R)$ where $\DiscreteStates$ and $\DiscreteActions$ are defined as in Definition \ref{def:mdp},  $\AggregatedTransitionProbability :\DiscreteStates\times\DiscreteActions\rightrightarrows \SpaceofDistributions(\DiscreteStates)$ is an \emph{uncertain transition probability function} such that for all $s$, $s'$ and $a$ it holds that $\AggregatedTransitionProbability(s,a)(s') \subseteq [0,1]$, and where $R :\DiscreteStates\rightarrow\Reals$ is a reward function. 
\end{definition}

\smallskip 
%Note that, since $\IntervalTransitionProbability(\StateAbs,\ActionAbs)\subseteq\SpaceofDistributions(\DiscreteStates)$, for all $\TransitionProbability(\StateAbs,\ActionAbs)\in\IntervalTransitionProbability(s,a)$ it holds that $\sum_{\StateAbs'\in\DiscreteStates}\TransitionProbability(\StateAbs,\ActionAbs)(\StateAbs') = 1$ and, hence, for all $\StateAbs$ and $\StateAbs'$ it holds that $\sum_{\StateAbs'\in \DiscreteStates} \min \IntervalTransitionProbability(\StateAbs,\ActionAbs)(\StateAbs')\leq 1 \leq \sum_{\StateAbs'\in \DiscreteStates} \max \IntervalTransitionProbability(\StateAbs,\ActionAbs)(\DiscreteStates')$.
%
\smallskip 

While RMDPs generalize iMDPs, both can be thought of as collections of MDPs each represented by an instance of a transition probability function $\TransitionProbability\in\AggregatedTransitionProbability$ or $\TransitionProbability\in\IntervalTransitionProbability$ respectively. For brevity, we present several useful notions for iMDPs; analogous observations hold for RMDPs. 

A deterministic time-varying policy for an iMDP is a function $\pi:\DiscreteStates\times\Naturals_0\rightarrow\DiscreteActions$, with $\pi\in\Pi_{M_{\updownarrow}}$ being the admissible policy space \cite{baier2008principles}.
A reach-avoid specification for an iMDP $\varphi'^K_{s_0}$ given a goal set $\DiscreteStates_G\subset\DiscreteStates$ and unsafe set $\DiscreteStates_U\subset\DiscreteStates$ is defined analogously to a specification for a SDE. Similarly, we denote the probability of satisfying a reach-avoid specification given a policy $\pi$ and a fixed transition probability function $\TransitionProbability\in\IntervalTransitionProbability$ as $\ProbabilityMeasure^\pi_\TransitionProbability\{\varphi'^K_{s_0}\}$. An optimal policy $\pi^*\in\Pi_{M_{\updownarrow}}$ for the iMDP maximises the worst-case probability of satisfying the specification with respect to all the possible transition probability functions coherent with the iMDP. Formally,
\begin{equation}\label{eq:opt-policy-imdp}
    \pi^*\in\argmax_{\pi\in\Pi_{M_{\updownarrow}}}\min_{\TransitionProbability\in\IntervalTransitionProbability}\ProbabilityMeasure^{\pi}_\TransitionProbability\{\varphi'^K_{s_0}\}.
\end{equation}
\begin{remark}\label{rem:reward-for-spec}
    Provably, the probability of satisfaction of a reach-avoid specification on an MDP can be equivalently expressed by computing the value function for a reward function defined as $R(\StateAbs)=1$ for all $\StateAbs\in\DiscreteStates_G$ and by making all states $s\in\DiscreteStates_U$ absorbing, i.e. $\TransitionProbability(\StateAbs,\ActionAbs)(\StateAbs)=1$, see \cite{baier2008principles}.
\end{remark} 
% Analogous observations hold for AMDPs.

% removed to save space -- Andrea
% We conclude this section by giving an overview of the rest of the letter. In Section \ref{sec:abstraction} we recall the construction scheme proposed by \cite{BRAPPSJ23} to abstract a DTSDS with additive noise to an MDP and introduce a shortcoming of such a procedure. In Section \ref{sec:aggregation} we propose a new scheme to abstract a DTSDS with additive noise to an AMDP where the aggregated transition probability function comprises a finite number of transition probability functions. Both schemes are viable only under the assumption that the distribution of the noise is known. In Section \ref{sec:pac-intervals} we relax such assumption, and instead, we rely on noise samples and scenario theory to construct an AMDP where the aggregated transition probability function comprises an infinite number of transition probability functions. For computational reasons, we embed the newly obtained AMDP to an iMDP. We conclude this work by presenting some insightful numerical examples in Section \ref{sec:experiments}.

%
\section{Finite-State Abstraction} \label{sec:abstraction}
Next, we describe the components required to construct a finite-state abstraction of a SDE: discretization of the state space,  transitions among the abstract states, and the evaluation of the probability associated with each transition.

\subsection{State Space Discretization}
\begin{figure}[h]
     \centering
     \begin{subfigure}[b]{0.45\linewidth}
         \centering
         \includegraphics[width=0.85\textwidth]{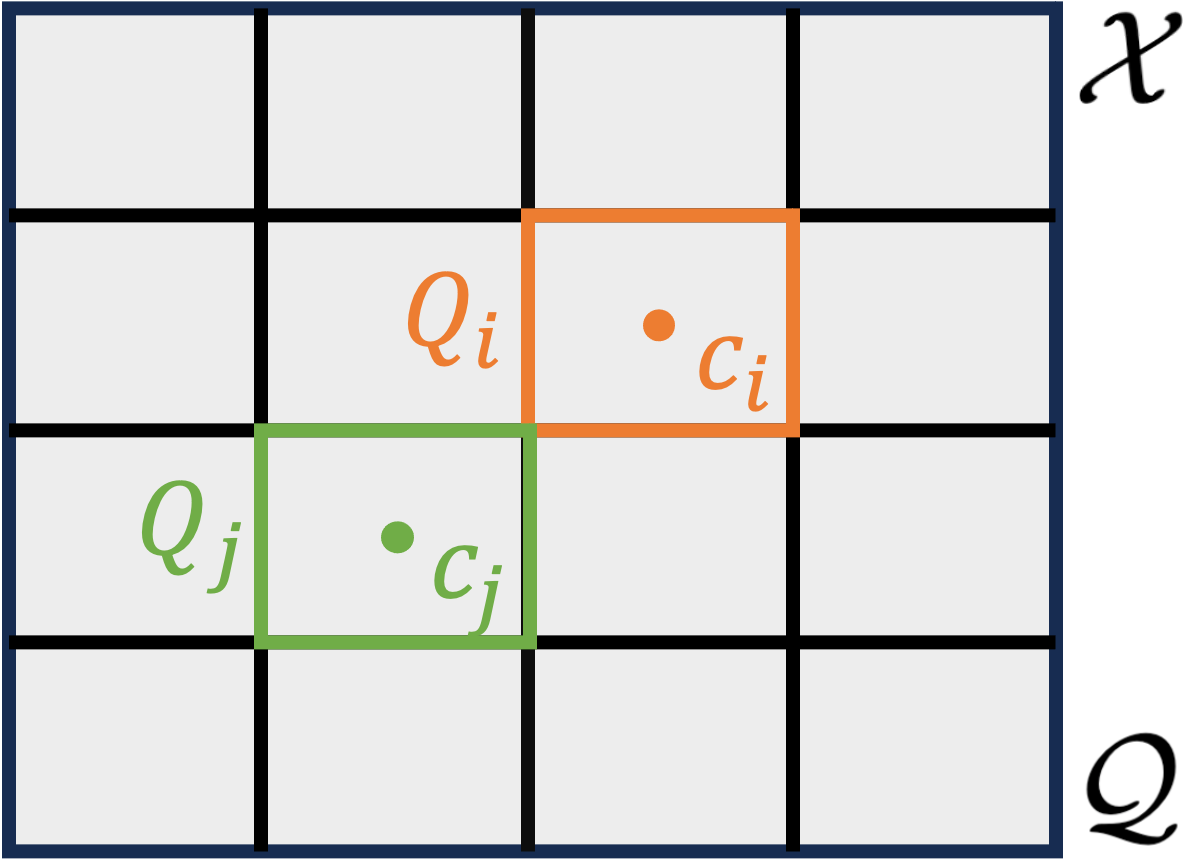}
         \caption{}
         \label{fig:domain-partition}
     \end{subfigure}
     \hfill
     \begin{subfigure}[b]{0.45\linewidth}
         \centering
         \includegraphics[width=0.85\textwidth]{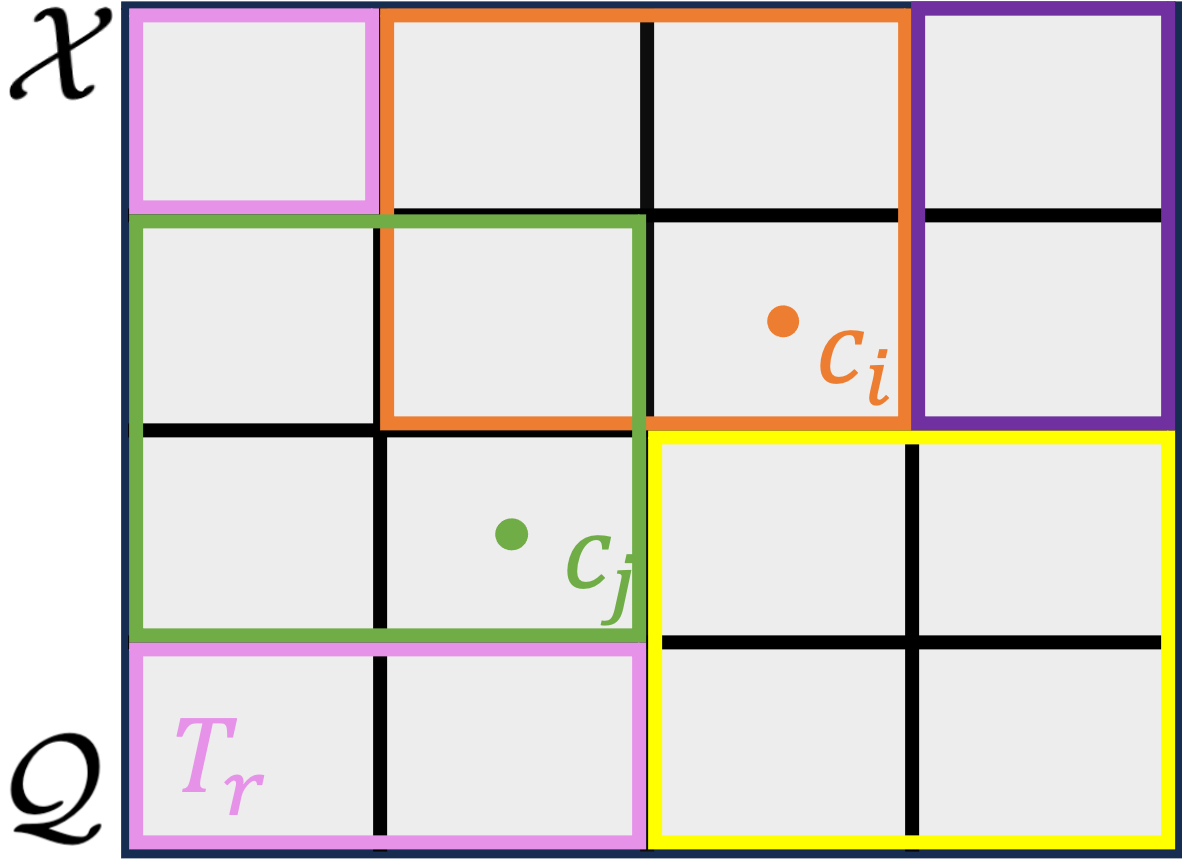}
         \caption{}
         \label{fig:target-sets}
     \end{subfigure}
        \caption{(a) Partition $\Partition$ of the domain of interest $\Domain$, where $\SetInPartition_i$ and $\SetInPartition_j$ are two elements of the partition, $c_i$ and $c_j$ are the respective reference points. (b) Cover of target sets $\TargetSets$ of the partition $\Partition$. Each color represents a different target set.}
        \label{fig:partition-&-cover}
\end{figure}

%
%\licio{If we are using this inverse mapping notation, do we need the $\mathrm{Pre}$ notation used in the sequel? Should we stick to one of it? Or am I missing something here?}
%\licio{As Alessandro suggested, we could frame this as relations between the corresponding space. The partition will be associated with equivalent relations; otherwise, we would have only a cover.}
Let $\Partition=\PartitionSetCollection$ be a partition of $\Domain\subset\Reals^\DimState$ such that every $\SetInPartition$ is an $\DimState$-dimensional convex polytope and let $\SetInPartition_0 = \text{cl}(\Reals^\DimState\setminus \Domain)$, i.e. the closure of $\Reals^\DimState\setminus \mathcal{X}$, be a so-called \emph{absorbing region}. 
We define an abstract state for each element of $\PartitionSetCollection$, yielding a set of $\SizePartition+1$ discrete states $\DiscreteStates = \{\StateAbs_i\}_{i = 0}^\SizePartition$. We define the relation $\PartitionRelation\subseteq\Reals^\DimState\times \DiscreteStates$ where $(\StateTransSys,\StateAbs_i)\in\PartitionRelation$ if and only if $\StateTransSys\in \SetInPartition_i$ for which we use the following notation $\PartitionRelation(\StateTransSys) := \{\StateAbs:(\StateTransSys,\StateAbs)\in \PartitionRelation\}$ and $\PartitionRelation^{-1}(\StateAbs_i):=\{\StateTransSys : (\StateTransSys,\StateAbs_i)\in\PartitionRelation\}=\SetInPartition_i$. %We assign to every abstract state $\StateAbs_i$ a unique arbitrary reference state $\ReferencePoint_i\in\SetInPartition_i$ by defining a bijective map $\MapToReference:\DiscreteStates\rightarrow\Domain$, with $\PartitionRelation(\ReferencePoint_i)=\StateAbs_i$ and $\MapToReference(\StateAbs_i)=\ReferencePoint_i$. 
Given a finite collection of \emph{reference} points in $\Domain$ denoted by $\{\ReferencePoint_i\}_{i=1}^\SizePartition$ such that $\PartitionRelation(\ReferencePoint_i)=\StateAbs_i$, we define a bijective map $\MapToReference:\DiscreteStates\rightarrow\{\ReferencePoint_i\}_{i=1}^\SizePartition$ as $\MapToReference(\StateAbs_i) = \ReferencePoint_i$. In our construction, the collection of points $\{\ReferencePoint_i\}_{i=1}^N$ is defined to be the centres of each element of the partition, as shown, for instance, in Figure \ref{fig:domain-partition}.
\begin{remark}
    Without loss of generality let the goal set $\Domain_G$ and unsafe set $\Domain_U$ be aligned with the partition, i.e. they can be equivalently expressed as a union of elements of $\Partition$; this allows to translate a specification on the concrete system $\varphi^K_{x_0}$ to a corresponding specification on a Markov model $\varphi'^K_{s_0}$.
\end{remark}

\subsection{Actions} \label{subsec:actions}
Below we define actions linking a single abstract state to (possibly) a \emph{set} of abstract states, named target set. 
Let $\TargetSets =\TargetSetCollection$ be a finite collection of \emph{target sets} covering $\Partition$, in particular $\TargetSets \subseteq \Powerset{\Partition}$, see Figure \ref{fig:target-sets}. 
Further, for every $i=1,...,\SizeTargetSets$,  
let $\ReferencePointCluster_i$ be the set of reference states associated with $\TargetSet_i$, that is $C_i = \bigcup_{\SetInPartition\in\TargetSet_i}\{\MapToReference(\PartitionRelation(x)) : \StateTransSys\in\SetInPartition\}$.

The collection of target sets defines the set of labelled abstract actions $\DiscreteActions=\{\ActionAbs_r : r=1,...,\SizeTargetSets\}$, where $\ActionAbs_r$ is associated with the target set $\TargetSet_r$ as shown below. We construct the set of enabled actions at $\StateAbs_i$ as follows. 

Action $\ActionAbs_r$ is enabled at state $\StateAbs_i$ if for every state $\StateTransSys_k\in\PartitionRelation^{-1}(\StateAbs_i)$ there exists a control input $\InputTransSys_k$ such that the next state under nominal dynamics \eqref{eq:nominal-dyn} belongs to the set of reference points associated with $\TargetSet_r$, or, in other words, $\hat{\StateTransSys}_{k+1} \in \ReferencePointCluster_r$.
Formally, we define the (nominal) \emph{backward reachable set} of a point $\StateTransSys'\in\Domain$ as
\begin{equation}\label{eq:pre-point}
    \text{Pre}(\StateTransSys'):=\{\StateTransSys\in\Domain: \exists u \in\InputSetTransSys, \Flow(\StateTransSys,\InputTransSys) = \StateTransSys'\}.
\end{equation}
We slightly abuse notation and define analogously the (nominal) \emph{backward reachable set} of a set $\ReferencePointCluster_r\subset\Domain$ as 
\begin{equation}\label{eq:pre-as-union}
    \text{Pre}(\ReferencePointCluster_r):=\bigcup_{c_j\in\ReferencePointCluster_r}\text{Pre}(c_j). 
\end{equation}
We require that backward reachable sets of reference points, or a union of those, can contain regions $\SetInPartition_i$. Thus, we lay the following assumption.
\begin{assumption}\label{ass:pre-has-volume}
    The backward reachable set of any reference point has a non-empty interior.
\end{assumption}
When the nominal dynamics is linear $\Flow(x_k,u_k) = Ax_k + Bu_k$, Assumption \ref{ass:pre-has-volume} requires that the matrix $B$ is full rank.
%\licio{
%Performing the previous reasoning for each element $q \in \Partition$, we obtain a collection of subsets of $\Partition$ which we denote by $\TargetSetCollection$, and a collection of discrete actions that we denote as $\DiscreteActions$, containing an action $a_i$ for each $\TargetSet_i$, $i = 1, \ldots, M$. In the sequel, we also use the notation
%\begin{equation}
%    \DiscreteActions(\StateAbs) = \{a \in \DiscreteActions: \PartitionRelation^{-1}(\StateAbs) \subseteq \text{Pre}(\ReferencePointCluster) \},
%    \label{eq:enabled-actions}
%\end{equation}
%where $\text{Pre}(\ReferencePointCluster)$ represents the center of the cluster associated to the collection $\TargetSet \in \TargetSetCollection$ .
%}
%\licio{We need to make sure the notation I have used above is consistent with the rest of the paper. My suggestion would be to order the finite collection $\DiscreteActions$ with elements $a_r$. If you agree with my suggestion above, could you implement this, Rudi?}

%\licio{This is the standard backward reachable set computation. I would definitely try to cite some papers by people in control that have studied this problem before.}

\begin{figure}[h]
     \centering
     \begin{subfigure}[b]{0.45\linewidth}
         \centering
         \includegraphics[width=\textwidth]{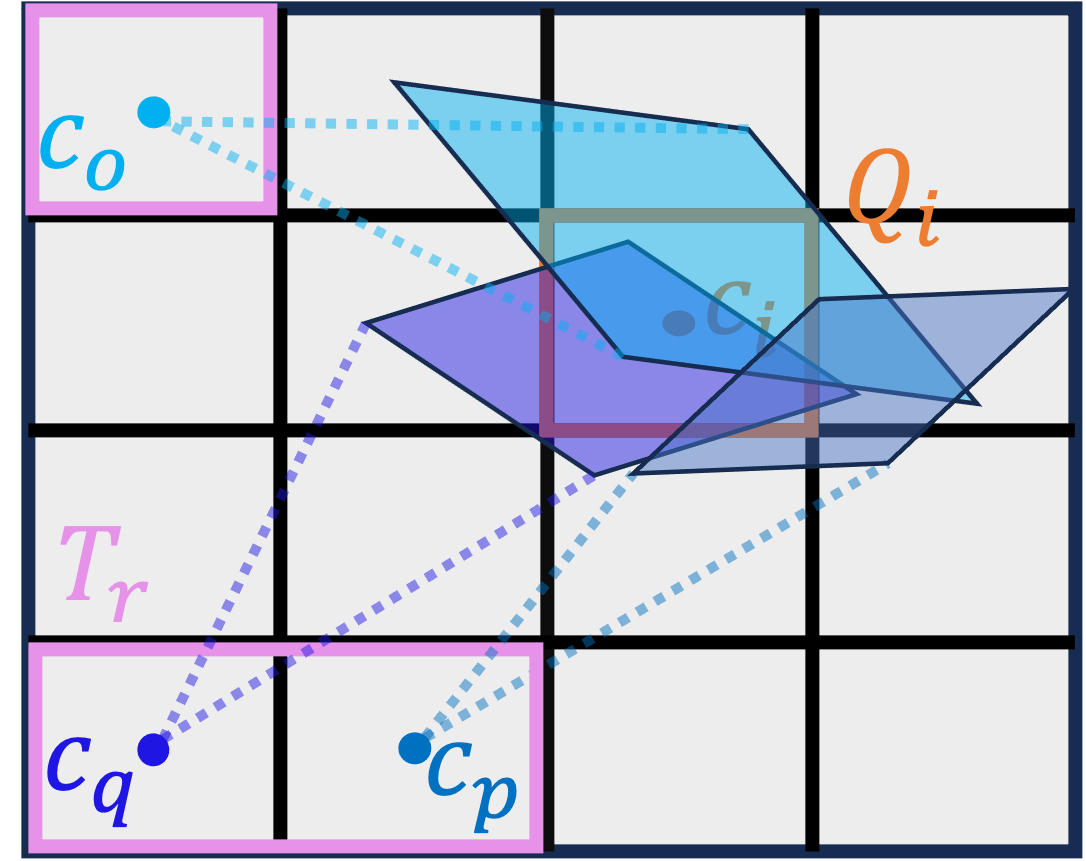}
         \caption{}
         \label{fig:nondeterminism-source}
     \end{subfigure}
     \hfill
     \begin{subfigure}[b]{0.33\linewidth}
         \centering
         \includegraphics[width=\textwidth]{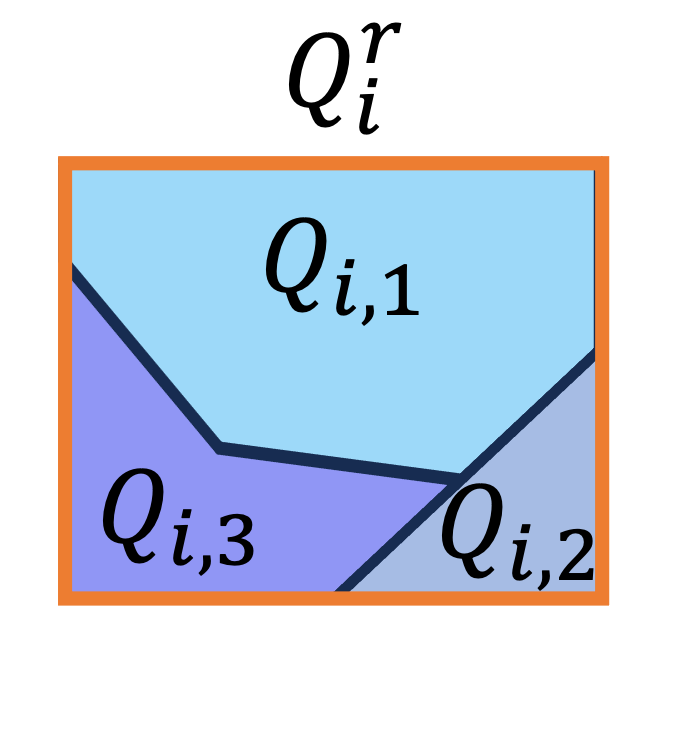}
         \caption{}
         \label{fig:subpartition}
     \end{subfigure}
        \caption{(a) $\text{Pre}(\ReferencePointCluster_r)$ represented as the union of $\text{Pre}(\ReferencePoint_o)$, $\text{Pre}(\ReferencePoint_p)$, and $\text{Pre}(\ReferencePoint_q)$; $\ActionAbs_r\in\DiscreteActions(\StateAbs_i)$ (b) Assuming the ordering $o < p < q$, the partition $\SetInPartition_i^r$ induced on $\SetInPartition_i$ by $\ReferencePointCluster_{r}$.}
        \label{fig:multi-target-cover}
\end{figure}

Action $\ActionAbs_r$ is enabled at state $\StateAbs_i$ iff $\PartitionRelation^{-1}(\StateAbs_i)$ is contained in $\text{Pre}(\ReferencePointCluster_r)$, as shown in Figure \ref{fig:nondeterminism-source}:
\begin{equation}\label{eq:contained-in-pre}
    \PartitionRelation^{-1}(\StateAbs_i) \subseteq \text{Pre}(\ReferencePointCluster_r) \iff \ActionAbs_r\in\DiscreteActions(\StateAbs_i).
\end{equation}

%In other words, for every $\StateTransSys \in \SetInPartition_i$, i.e.  such that $R(\StateTransSys)=\StateAbs_i$, if $\ActionAbs_r\in\DiscreteActions(\StateAbs_i)$ then there exists $u$ such that $f(x,u)$ equals one of the reference states associated with $\TargetSet_{\ell}$.

In general for a continuous state such that $x\in \text{Pre}(\ReferencePointCluster_r)$ there may exist multiple control inputs leading to $\ReferencePointCluster_r$, that is the sets $\{\text{Pre}(\ReferencePoint_j)\}_{\ReferencePoint_j\in\ReferencePointCluster_r}$ need not be disjoint. To unambiguously define a feedback controller $\InputTransSys:\Domain\rightarrow\InputSetTransSys$ we exploit the ordering of the partitioning sets $\PartitionSetCollection$ to assign a unique control input driving the state to $\ReferencePointCluster_r$.

Let $\ReferencePoint^*:\Domain\times\DiscreteActions\rightarrow\Domain$ be a function mapping a continuous state and abstract action to a continuous reference point indexed by the lowest integer, that is
\begin{align}
\label{eq:ref-point-star}
    \ReferencePoint^*(\StateTransSys,\ActionAbs_r) := &\argmin_{\ReferencePoint_j\in \ReferencePointCluster_r} j
    % \\
     &\text{s.t.} \quad \StateTransSys\in\text{Pre}(\ReferencePoint_j). 
     % \nonumber
\end{align}
Let us define the unique control law\footnote{Uniqueness can be guaranteed by choosing the minimum norm control if necessary. Details are omitted for brevity.} $\InputTransSys^*:\Domain\times\DiscreteActions\rightarrow\InputSetTransSys$ as
\begin{equation}\label{eq:control-law}
    \InputTransSys^*(\StateTransSys,\ActionAbs_r) :=\{\InputTransSys : \Flow(\StateTransSys,\InputTransSys)=\ReferencePoint^*\}.
\end{equation}

% \licio{We should definitely emphasize this section in the paper, as it seems to contain our main results. A suggestion could be to add a remark contrasting this new discrete model with the ones used in the literature (my work with Thom, yours, and possibly others in the literature). If I am the reader, I want to be convinced at this stage about the usefulness of the model we are putting forward. And contrasting it with model that the reader may be familiar could add value to our work.

% I have noticed that this is done in section D below. But I would argue that it could be nicer to engage the reader about the importance of the model at this stage, in order not to loose their attention and interest before they arrive in Section D.}
For every $\ReferencePointCluster_r$, 
operation \eqref{eq:ref-point-star} naturally induces a partition on a set $\PartitionRelation^{-1}(s_i)$ satisfying \eqref{eq:contained-in-pre}, 
% and the resulting partition $\mathcal{Z}_i^\ell$ on $\SetInPartition_i$
defined as
\begin{equation}
\label{eq:sub-partition}
    \SetInPartition_i^{r}:=\{X\subseteq \SetInPartition_i : \forall \StateTransSys,\StateTransSys'\in X,  \ReferencePoint^*(\StateTransSys,\ActionAbs_r)=\ReferencePoint^*(\StateTransSys',\ActionAbs_r)\}.
\end{equation}
In other words, the partition $\SetInPartition_i^{r}=\{\SetInPartition_{i,\ell}\}_{\ell=1}^{\SubPartitionSize_r}$ is a collection of $\SubPartitionSize_r$ non-overlapping regions  of $\SetInPartition_i$ defined by the points sharing the same next state under nominal dynamics under control law $\InputTransSys^*$, see Figure \ref{fig:subpartition}.

\subsection{Transition Probabilities}\label{subsec:old_paper}
\begin{figure}[h]
     \centering
     \begin{subfigure}[b]{0.45\linewidth}
         \centering
         \includegraphics[width=\textwidth]{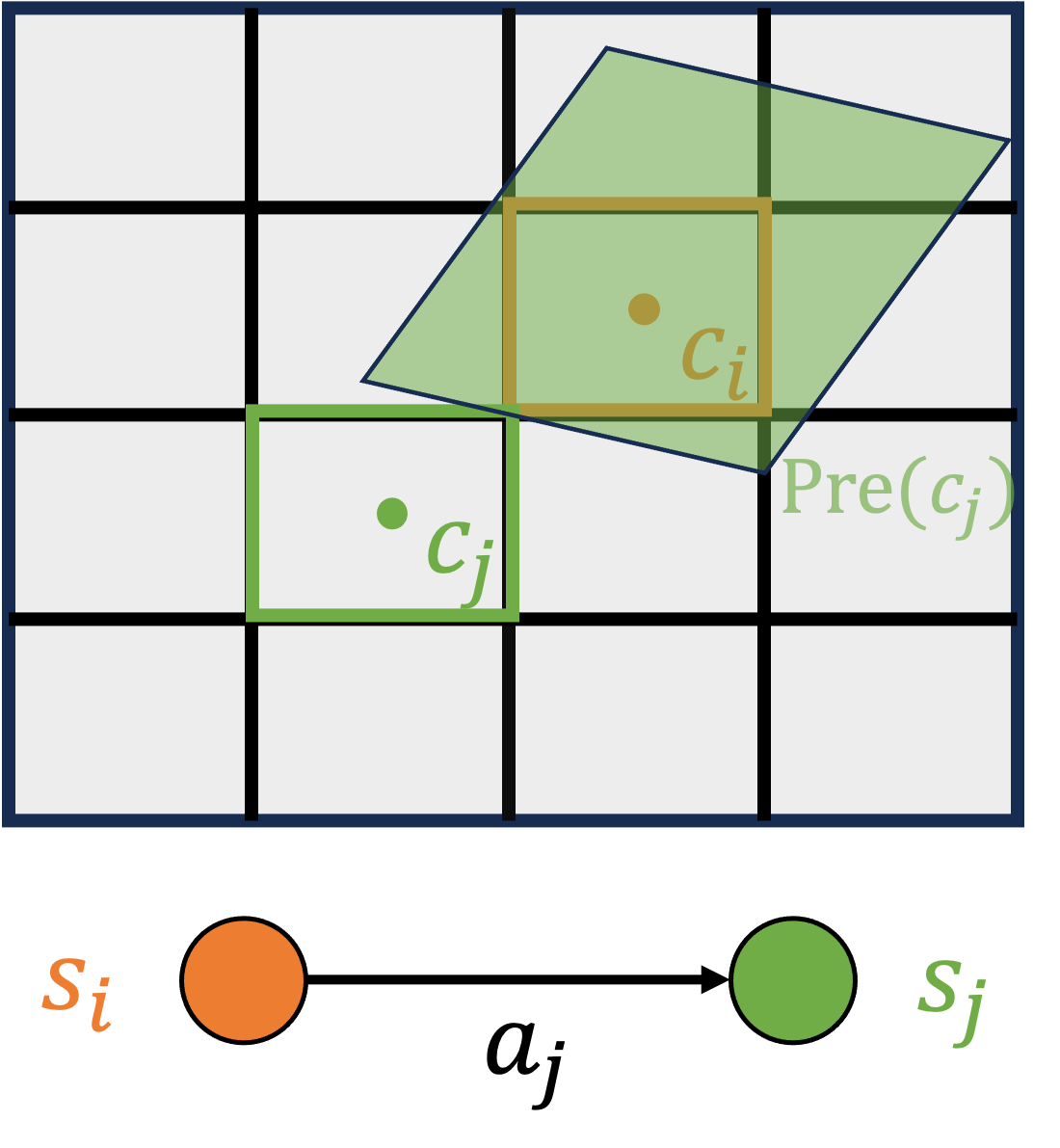}
         \caption{}
         \label{fig:old_paper_transitions}
     \end{subfigure}
     \hfill
     \begin{subfigure}[b]{0.44\linewidth}
         \centering
         \includegraphics[width=\textwidth]{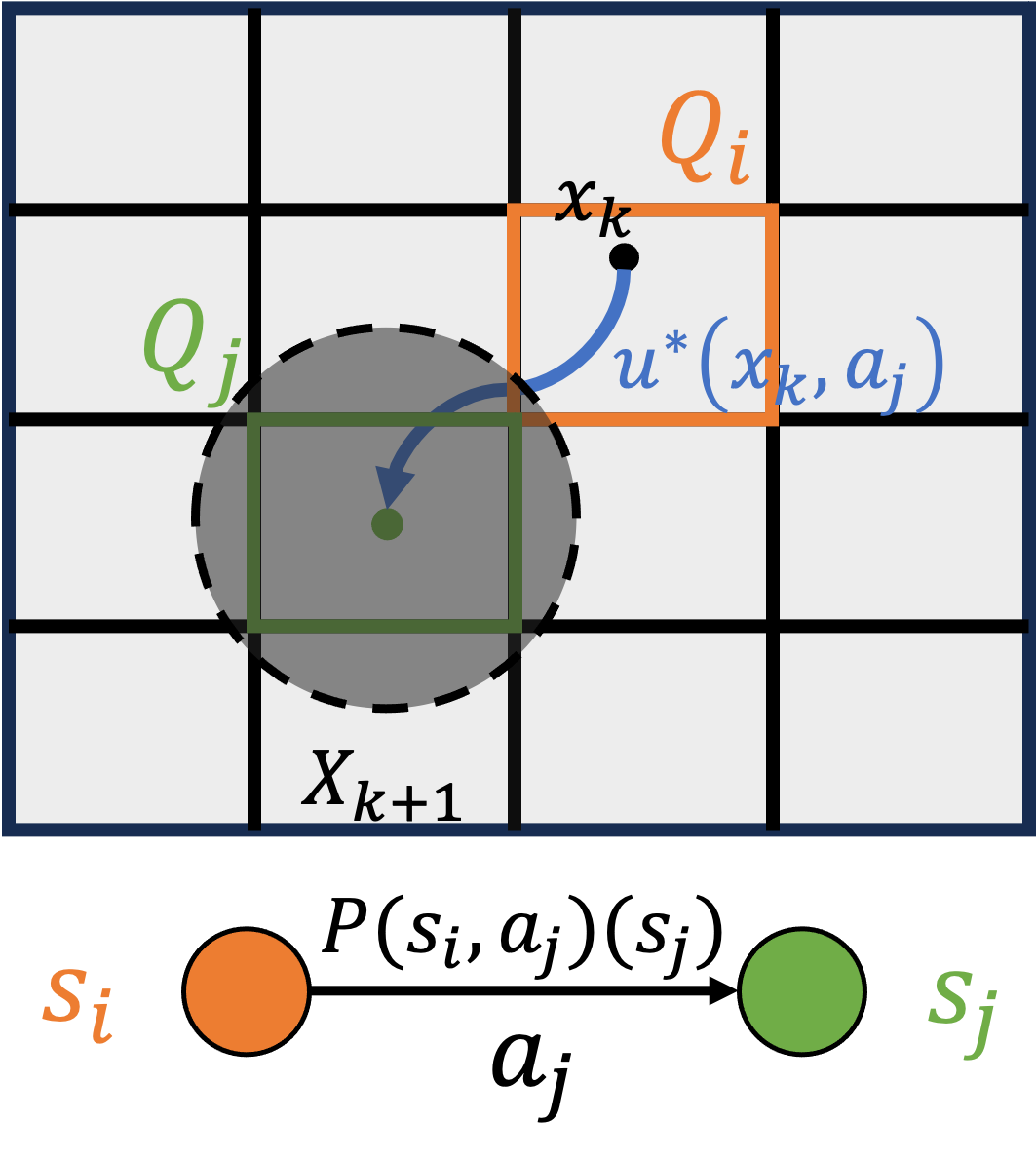}
         \caption{}
         \label{fig:old_paper_probability}
     \end{subfigure}
        \caption{(a) Consider $\TargetSet_j=\SetInPartition_j$. If action $\ActionAbs_j\in\DiscreteActions(\StateAbs_i)$ then $\text{Pre}(\ReferencePoint_j)\supseteq \SetInPartition_i$. (b) For every $\StateTransSys_k\in \SetInPartition_i$ there exists an input $\InputTransSys_k=\InputTransSys^*(\StateTransSys,\ActionAbs_j)$ driving the state to $\ReferencePoint_j$. The shaded area represents the support of $\StochasticKernel(d\StateTransSys_{k+1}|\StateTransSys_k,\InputTransSys_k)$.}
        \label{fig:old-paper}
\end{figure}
% \licio{Move this picture so that it does not appear as the first paragraph in this Section.}
%
In the remaining part of this section, we recall the construction scheme proposed by \cite{BRAPPSJ23} to abstract a SDE with additive noise to an MDP and introduce a shortcoming of such a procedure.
Suppose that the collection of target sets $\TargetSets$ coincides with the partition $\Partition$, more precisely, $\TargetSet_r=\{\SetInPartition_r\}$ for every $r=1,...,\SizePartition$. 
In this tailored setting we can simplify our discussion: 
% for every $\ActionAbs_j\in\DiscreteActions(\StateAbs_i)$ 
every set $\ReferencePointCluster_r$ contains a single element, namely $\ReferencePoint_r$,
hence action $\ActionAbs_r$ is enabled in the abstract state $\StateAbs_i$ if and only if $\PartitionRelation^{-1}(\StateAbs_i)\subseteq\text{Pre}(\ReferencePointCluster_r)=\text{Pre}(\ReferencePoint_r)$. 
Similarly, $\SetInPartition_i^r$ is the trivial partition and contains a single element, namely $\SetInPartition_i^r = \{ \SetInPartition_i \}$ -- cfr. \eqref{eq:sub-partition} -- as depicted in Figure~\ref{fig:old_paper_transitions}. 
A finite-state abstraction that describes this framework is an MDP $M = (\DiscreteStates,\DiscreteActions,\TransitionProbability,R)$, where given $\StateTransSys_k\in\PartitionRelation^{-1}(\StateAbs_i)$, an abstract action $\ActionAbs_j\in\DiscreteActions(\StateAbs_i),$ and $\InputTransSys_k = \InputTransSys^*(\StateTransSys_k,\ActionAbs_j)$ the probability of transitioning to the abstract state $s_j$ can be computed as: 
\begin{equation}\label{eq:transition-prob}
    \TransitionProbability(\StateAbs_i,\ActionAbs_j)(\StateAbs_j) := \int_{\PartitionRelation^{-1}(\StateAbs_j)}\StochasticKernel(d\StateTransSys_{k+1} | \StateTransSys_k, \InputTransSys_k)    . 
\end{equation}
Due to the noise being additive \eqref{eq:add-noise} and given the control law $\InputTransSys^*$ we can express \eqref{eq:transition-prob} as  
\begin{equation}\label{eq:mdp-transition-prob}
    \TransitionProbability(\StateAbs_i,\ActionAbs_j)(\StateAbs_j)
    = \ProbabilityMeasure\{\Sample\in\SampleSet : \ReferencePoint_j + \RVProcessNoise_k(\Sample)\in\PartitionRelation^{-1}(\StateAbs_j)\}, 
\end{equation}
where we have used the fact that under nominal dynamics $\Flow(\StateTransSys_k,\InputTransSys^*(\StateTransSys,\ActionAbs_j))=\ReferencePoint_j$. This situation is depicted in Fig.~\ref{fig:old_paper_probability}.

% \licio{This example is nice! I was thinking whether the paper could start with it, rather than it appearing only here. We could start the paper with a brief description of the example and its main shortcoming. Then, once we are introducing the model in Section $B$ above, we could retrieve the discussion to better motivate our work.}
\subsection{Shortcomings and Motivating Example}

One shortcoming of this approach is that it may lead to a significant \emph{under-approximation} of the dynamics of the concrete system. 
Formally expressed in \eqref{eq:contained-in-pre}, if $\SetInPartition_i$ is not entirely contained in $\Pre(\ReferencePoint_j)$, $\ActionAbs_j$ is not enabled for $\StateAbs_i$. 
As such, 
one may have to exclude a large set of actions if the dynamics are not well aligned with the chosen partition.
\begin{figure}[h]
    \centering
    \includegraphics[width=0.45\linewidth]{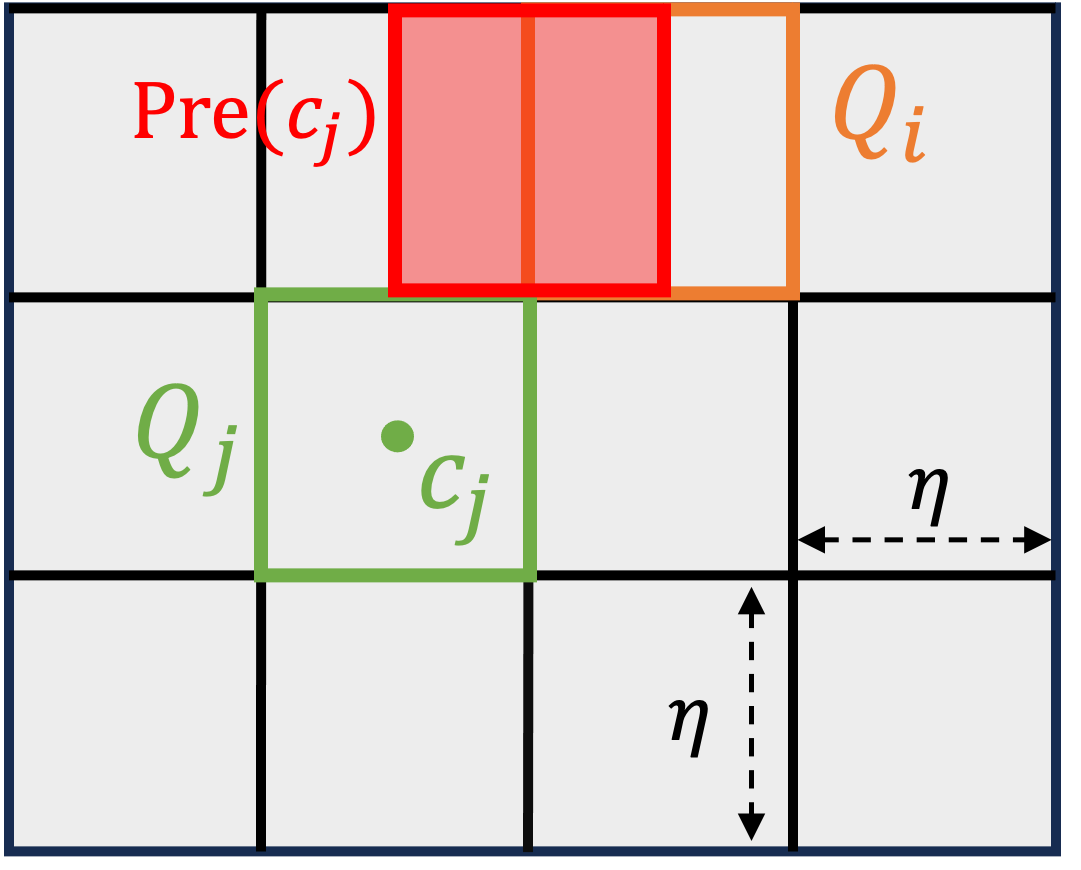}
    \caption{The dynamics are misaligned with the partition.}
    \label{fig:motivating_example}
\end{figure}

\begin{example} \label{exa:motivation}
Consider the SDE given by 
\begin{equation}
\label{eq:example-dynamics}
    \RVStateTransSys_{k+1} = \RVStateTransSys_k - \InputTransSys_k + W_k, 
\end{equation}
where $\RVStateTransSys_k \in \mathbb{R}^2$, 
$W_k$ is a random variable taking values in $\mathbb{R}^2$, $\InputTransSys_k \in \mathcal{U} = [0,\eta]\times[\eta/2, 3\eta/2]\subset\mathbb{R}^2$ for some $\eta>0$. Let the partition $\Partition$ of $\mathcal{X}$ be a uniform grid such that each set $\SetInPartition_i$ is a $\eta$ by $\eta$ box. Suppose that $\TargetSets = \Partition$. Consider a reference state $c_j$, the center of the box $S_j$, and let us examine $\text{Pre}(T_j)=\text{Pre}(c_j)$: by inverting the nominal dynamics we can characterize such set as
\begin{equation*}
    \text{Pre}(c_j) = \{x: \exists u \in \mathcal{U}, c_j + u = x \}, 
\end{equation*}
which represents a copy of $S_j$ with its center shifted by $[\eta/2,\eta]$, as represented in Figure \ref{fig:motivating_example}. Consequently, considering only one partition for each transition  
% the construction presented in Section \ref{subsec:old_paper} 
leads to an empty MDP: i.e., all abstract states $\StateAbs_i$ have an empty action set. 
By considering a different target set, for instance partition $Q_j$ and its adjacent partition, it is easy to see that  $\SetInPartition_i$ is entirely contained in the union of the two $\Pre$ sets: this observation is the motivation to the following discussion. 
\end{example}

\section{Uncertain Transition Probabilities}\label{sec:aggregation}

In contrast to the previous discussion, let us consider now consider a general cover $\TargetSets$ of the partition $\Partition$. 
%, such that \emph{every} single partition $\SetInPartition_i$ is entirely contained in at least one $\Pre(\TargetSet_{\ell})$, $\ell = 1, \ldots, M$. 
%Such cover can be selected by carefully analysing the dynamics of \eqref{eq:add-noise}; for ease of presentation, we provide more details regarding this computation in the following section. 
%
, i.e. at least one out of the $\SizeTargetSets$ target sets, say $\TargetSet_r$, contains more than one element of $\Partition$; equivalently, $\ReferencePointCluster_r$ contains more than one reference state.  
Let $\SetInPartition_i \subseteq \text{Pre}(\ReferencePointCluster_r)$, as depicted in Figure \ref{fig:nondeterminism-source}, and consider the non-trivial partition $\SetInPartition_i^r=\{\SetInPartition_{i,\ell}\}_{\ell=1}^{\SubPartitionSize_r}$ induced on $\SetInPartition_i$ by \eqref{eq:ref-point-star} and described by \eqref{eq:sub-partition}. We know that for every $\ell=1,...,\SubPartitionSize_r$ and for every $\StateTransSys \in \SetInPartition_{i,\ell}$ there exists a control law that can drive the state to the center of one of the reference points in $\ReferencePointCluster_r$, namely $\InputTransSys^*(x,\ActionAbs_r)$.

%Consider now the transition $\SetInPartition_i \to T_{\ell}$, where $\TargetSet_{\ell}$ contains $K$ partitions, i.e. $T_{\ell} = \{ S_{j1}, S_{j2}, \ldots S_{jK}\}$, 
%and let us recall the partitioning of $\SetInPartition_i$ as  $\mathcal{Z}_i^\ell = \{ Z_{i, m}^\ell\}_{m=1}^K$ (see \eqref{eq:sub-partition}). 
%Each portion $Z_i{,m}^\ell$ contains\footnotemark \   the continuous states $x \in \SetInPartition_i$ for which there exists a control input that drives the dynamics to the reference point of $S_{jm} \in \TargetSet_{\ell}$. 
%\footnotetext{Not necessarily all. -- to be quickly recalled -- }

%Each individual transition $Z_{i,j}^\ell \to S_{j}$ is governed by a transition probability distribution, that, similarly to \eqref{eq:mdp-transition-prob}, depends on the transition kernel $p_X$. 
%
Naturally, in this new setting, it is not possible to describe the transition from $\StateAbs_i$ under action $\ActionAbs_r$ to a future abstract state $\StateAbs_j = \PartitionRelation(\ReferencePoint)$ for $ \ReferencePoint\in \TargetSet_r$ by a \emph{single} transition probability function as in \eqref{eq:mdp-transition-prob}, but rather by a \emph{set} of transition probability functions. 
%
% It is not possible to capture a state-action-state transition with a single transition probability function. This is because, in general, given $\StateAbs_i$ and $\ActionAbs_\ell\in\DiscreteActions_\ell$ it is not possible to associate a single reachable reference state $c_\ell$ such that $f(x,u(x,c_\ell)) = c_\ell$ for all $x\in R^{-1}(\SetInPartition_i)$. Hence, the probability of reaching $\StateAbs_j$ from state $\StateAbs_i$ under action $\ActionAbs_\ell$ will depend on the continuous state related to $\StateAbs_i$ under consideration. See Figure \ref{fig:nondeterminism-source}.
%
Indeed, the probability of reaching $\StateAbs_j$ from $\StateAbs_i$ under action $\ActionAbs_r$ depends on the actual continuous state $\StateTransSys \in \SetInPartition_i$ from which the transition takes place. 

\begin{figure}[h]
     \centering
     \begin{subfigure}[b]{0.45\linewidth}
         \centering
         \includegraphics[width=\textwidth]{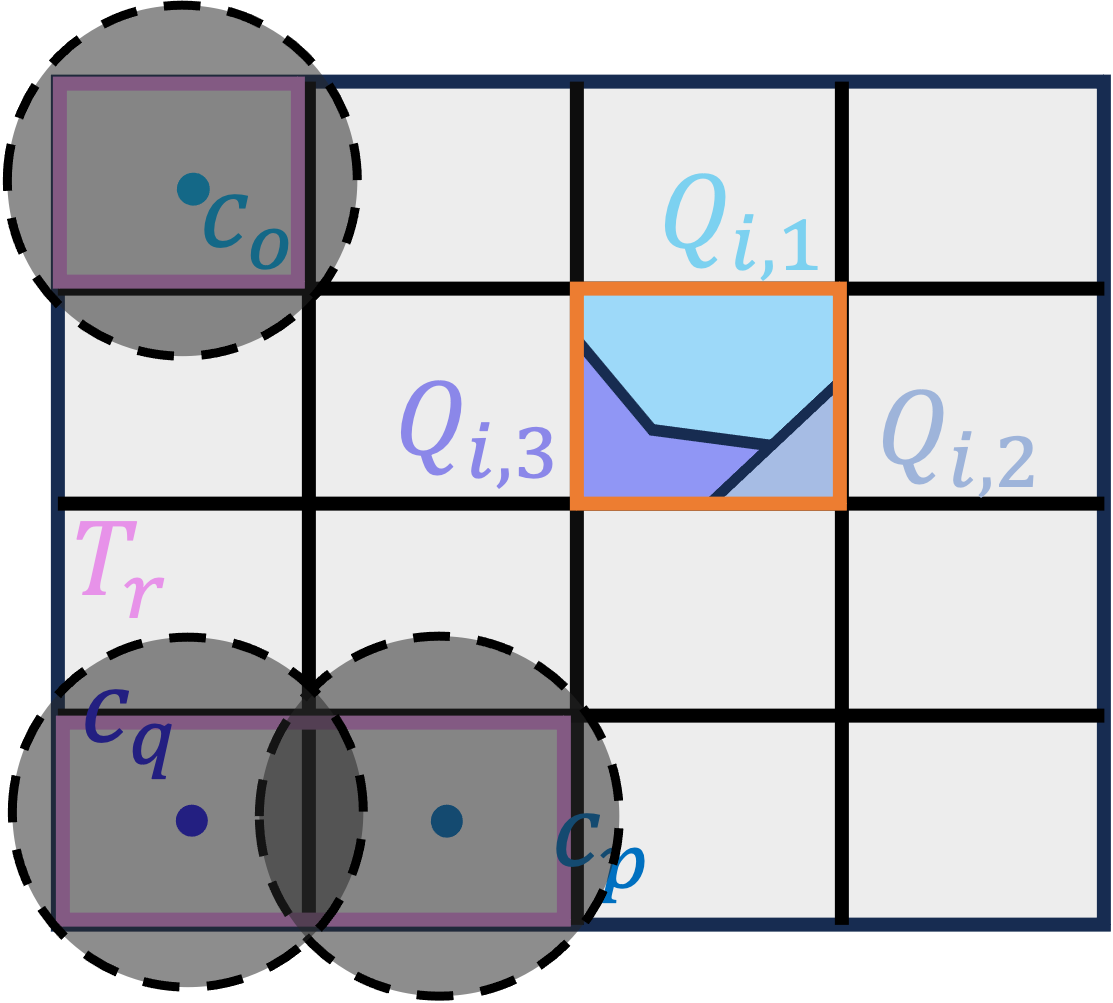}
         \caption{}
         \label{fig:multi_target}
     \end{subfigure}
     \hfill
     \begin{subfigure}[b]{0.50\linewidth}
         \centering
         \includegraphics[width=\textwidth]{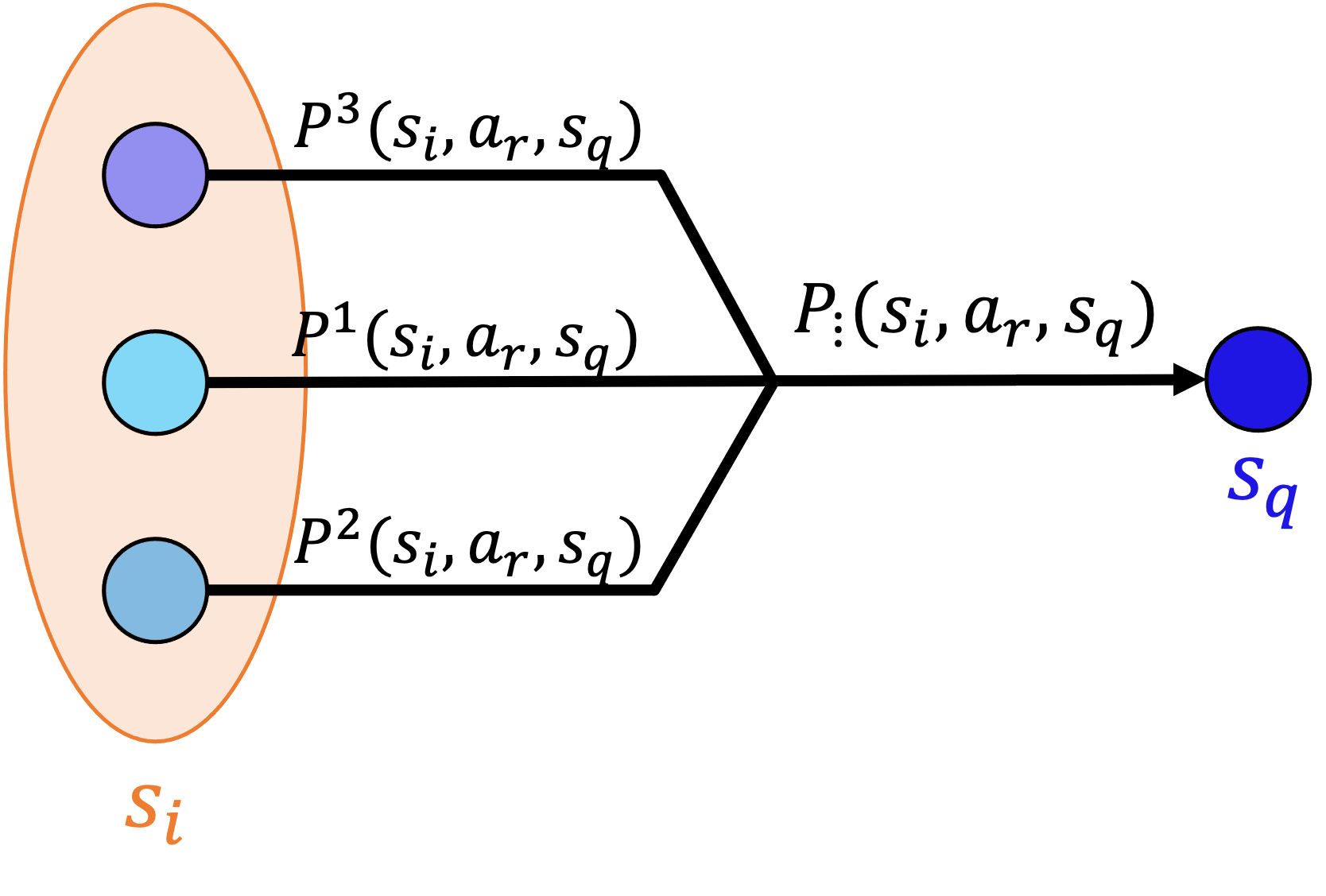}
         \caption{}
         \label{fig:aggregated-transition-probability}
     \end{subfigure}
        \caption{(a) Computation of $\TransitionProbability^\ell(\StateAbs_i,\ActionAbs_r)(\StateAbs_q)$ as per \eqref{eq:individual-transition-probability} for $\ell=1,2,3$. (b) Uncertain transition probability function from a state $\StateAbs_i$ to a state $\StateAbs_q$ under action $\ActionAbs_r$.}
        \label{fig:nondeterminism}
\end{figure}

In order to encompass this framework, 
we define an uncertain probability transition function $\AggregatedTransitionProbability$ which encapsulates all possible cases and captures the \emph{nondeterminism} introduced by clustering multiple reference points. Consider an abstract state $\StateAbs_i$, an action $\ActionAbs_r\in\DiscreteActions(\StateAbs_i)$, the partitioning $\SetInPartition_i^r$, and suppose that $\StateTransSys_k\in \SetInPartition_{i,\ell}$ for some $\ell\in\{1,...,\SubPartitionSize_r\}$: under the control input $\InputTransSys_k=\InputTransSys^*(\StateTransSys_k,\ActionAbs_r)$ the next state under nominal dynamics is the reference point $\ReferencePoint^*(\StateTransSys,\ActionAbs_r)\in\ReferencePointCluster_r$. 
Let us define 
\begin{equation}\label{eq:individual-transition-probability}
    \TransitionProbability^\ell(\StateAbs_i,\ActionAbs_r)(\StateAbs_j) := \int_{\PartitionRelation^{-1}(\StateAbs_j)}\StochasticKernel(d\StateTransSys_{k+1} | \StateTransSys_k, \InputTransSys_k)   ,
\end{equation}
By enumerating $\ell = 1, ..., \SubPartitionSize_r$ we obtain a set of transition probability functions which describes all cases, namely $\StateTransSys\in\SetInPartition_{i,1}, ...,\StateTransSys\in\SetInPartition_{i,\SubPartitionSize_r}$. Accordingly, we define the RMDP $M_{\fcolon}=(\DiscreteStates,\DiscreteActions,\AggregatedTransitionProbability,R_{\fcolon})$ where 
\begin{equation}
    \AggregatedTransitionProbability(\StateAbs_i,\ActionAbs_r)(\StateAbs_j) = \bigcup_{\ell=1}^{\SubPartitionSize_r} \TransitionProbability^\ell(\StateAbs_i,\ActionAbs_r)(\StateAbs_j).   
\end{equation}
This is shown graphically in Figure \ref{fig:multi_target} and Figure \ref{fig:aggregated-transition-probability}.

\begin{remark}
    % The AMDP setting provides the additional degree of freedom of choosing multiple target partitions. These can be selected by inspecting the backward reachable set of a single partition -- which provides an intuition on the number of needed partitions and their respective configuration -- and the same configuration can be applied for all states. 
    The target sets can be selected by inspection of the backward reachable set. An intuitive choice is to select adjacent cells, creating `neighborhoods' of increasing size. 
\end{remark}

%%%%%%%%%%%%%%%%%%%%%%%%%%%%%%%%%%%%%%%%%%%%%%%%%

\section{PAC Probability Intervals via Sampling}\label{sec:pac-intervals}

Computing \eqref{eq:individual-transition-probability} is possible only when the distribution of the additive noise $W_k$ is perfectly known. Additionally, even if it were known, computing explicitly the integral could be difficult or undesirable in certain cases. Instead we provide a lower and upper bound of $\TransitionProbability^\ell(\StateAbs_i,\ActionAbs_r)(\StateAbs_j)$ using the \emph{sampling-and-discarding} scenario approach proposed in \cite{campi2011sampling} and improved in \cite{romao2022exact}. In particular, we adopt the framework presented in \cite{BRAPPSJ23}, under the following necessary assumption
\begin{assumption}[Non-degeneracy]
    For every $k$, $\RVProcessNoise_k$ has a density with respect to the Lebesgue measure.
\end{assumption}

We summarize the results therein here. 
Let us collect a set of $\NumSamples\in\Naturals$ i.i.d. samples of $\RVProcessNoise_k$, denoted $\ProcessNoise_k^{(1)},...,\ProcessNoise_k^{(\NumSamples)}$ and define the quantities
\begin{align*}
    \NumSamples_{\StateAbs_j}^{\text{in}} &= |\{\ProcessNoise_k^{(i)}: \ProcessNoise_k^{(i)} + \ReferencePoint_j \in \SetInPartition_j\}|,\quad  \NumSamples_{\StateAbs_j}^{\text{out}} &= \NumSamples - \NumSamples_{\StateAbs_j}^{\text{in}}.
\end{align*}
In words, $\NumSamples_{\StateAbs_j}^{\text{in}}$ is the number of samples $\ProcessNoise_k^{(i)}$ which, when shifted by $\ReferencePoint_j$, fall within region $\SetInPartition_j$. 

\begin{theorem}(PAC probability intervals \cite[Theorem~1]{BRAPPSJ23})\label{thm:pac-prob-intervals}
    Given $\NumSamples$ samples of the noise $\RVProcessNoise_k$, compute $\NumSamples_{\StateAbs_j}^{\text{out}}$ and fix a confidence parameter $\beta$. It holds that
    \begin{equation*}
        \ProbabilityMeasure^\NumSamples\{{\MinProb_{j,\ell}}\leq \TransitionProbability^\ell(\StateAbs_i,\ActionAbs_r)(\StateAbs_j) \leq {\MaxProb_{j,\ell}}\} \geq 1-\beta,
    \end{equation*}
    where ${\MinProb_{j,\ell}}= 0$ if $\NumSamples_{\StateAbs_j}^{\text{out}}=\NumSamples$; otherwise ${\MinProb_{j,\ell}}$ is the solution of
    \begin{equation*}
        \frac{\beta}{2\NumSamples} = \sum_{i=0}^{\NumSamples_{\StateAbs_j}^{\text{out}}} {\NumSamples \choose i}(1-{\MinProb_{j,\ell}})^i{\MinProb_{j,\ell}}^{\NumSamples-i},
    \end{equation*}
    and ${\MaxProb_{j,\ell}}=1$ if $\NumSamples_{\StateAbs_j}^{\text{out}}=0$; otherwise ${\MaxProb_{j,\ell}}$ is the solution of
    \begin{equation*}
        \frac{\beta}{2\NumSamples} = 1-\sum_{i=0}^{\NumSamples_{\StateAbs_j}^{\text{out}}-1} {\NumSamples \choose i}(1-{\MaxProb_{j,\ell}})^i{\MaxProb_{j,\ell}}^{\NumSamples-i} . 
    \end{equation*}
\end{theorem}
Theorem \ref{thm:pac-prob-intervals} allows us to provide an upper and lower bound on the individual transition probabilities $\TransitionProbability^\ell(\StateAbs_i,\ActionAbs_r,\StateAbs_j)$.

We can then describe the resulting abstraction as a RMDP $M_{\fcolon}=(\DiscreteStates,\DiscreteActions,\AggregatedTransitionProbability,R_{\fcolon})$ where 
\begin{equation}\label{eq:underlying-ampd}
    \AggregatedTransitionProbability(\StateAbs_i,\ActionAbs_r)(\StateAbs_j) := \bigcup_{\ell=1}^{\SubPartitionSize_r} [{\MinProb_{j,\ell}}, {\MaxProb_{j,\ell}}]. 
\end{equation}

%\rudi{comment on the validity of all the intervals}
%\licio{Interestingly enough, we have done this computation and proved the validity of this argument in this paper: \href{https://link.springer.com/chapter/10.1007/978-3-031-43835-6_2}{here [equation $(15)$]}. The context is different as we were interested in studying Markov jumping linear systems. But I believe there is some strong connection with this point.}
%
In order to leverage existing algorithms for value iteration on iMDPs, following the approach in \cite{givan2000bounded,dean1997model}, we can embed (abstract) the resulting RMDP into an iMDP $M_\updownarrow=(\DiscreteStates,\DiscreteActions,\IntervalTransitionProbability,R)$ where the uncertain transition probability from $\StateAbs_i$ to state $\StateAbs_j$ under action $\ActionAbs_r$ is defined as
\begin{equation}\label{eq:embedding-to-imdp}
    \IntervalTransitionProbability(\StateAbs_i,\ActionAbs_r)(\StateAbs_j) = [\MinProb_j,\MaxProb_j], 
\end{equation}
where $\MinProb_j = \min\AggregatedTransitionProbability(\StateAbs_i,\ActionAbs_r)(\StateAbs_j)$ and $\MaxProb_j = \max\AggregatedTransitionProbability(\StateAbs_i,\ActionAbs_r)(\StateAbs_j)$.

It is obvious from \eqref{eq:underlying-ampd} and \eqref{eq:embedding-to-imdp} that the collection of MDPs described by $M_{\fcolon}$ is a subset of the MDPs described by $M_{\updownarrow}$. Indeed if $\TransitionProbability\in\AggregatedTransitionProbability$ it implies that $\TransitionProbability\in\IntervalTransitionProbability$. Let $\pi^*$ denote the optimal policy for $M_{\updownarrow}$. It follows that
\begin{equation*}
\min_{\TransitionProbability\in\IntervalTransitionProbability}\ProbabilityMeasure^{\pi^*}_\TransitionProbability\{\varphi'^K_{s_0}\}\leq\min_{\TransitionProbability\in\AggregatedTransitionProbability}\ProbabilityMeasure^{\pi^*}_\TransitionProbability\{\varphi'^K_{s_0}\}. 
\end{equation*}
This embedding allows us to employ existing tools to obtain a policy with a valid lower bound on the probability of satisfaction of the reach-avoid property for the RMDP. Note that we compute the optimal policy $\pi^*$ with respect to the iMDP, which in general differs from the optimal policy with respect to the RMDP. We leave this for future work.

%Recalling Remark \ref{rem:reward-for-spec}, we leverage Assumption \ref{ass:aligned-goal-set} to translate a specification on the DTSDS $\varphi^K_{x_0}$ to a corresponding specification on an iMDP $\varphi'^K_{s_0}$.
%\begin{assumption}\label{ass:aligned-goal-set}
%    The goal set $\Domain_G$ and unsafe set $\Domain_U$ can be equivalently expressed as a union of elements of $\Partition$.
%\end{assumption}

We conclude this section by enunciating the following theorem connecting the probability of satisfaction of the reach-avoid property on the iMDP given an optimal policy with the probability of satisfaction of the reach-avoid property on the underlying dynamical system by refining the policy to a feedback controller. The proof follows the rationale in \cite[Theorem~2]{BRAJ23a}, and is omitted for brevity.

\begin{theorem}[Adapted from \cite{BRAJ23a}]
    Let $\pi^*$ denote the optimal policy \eqref{eq:opt-policy-imdp} for the iMDP obtained according to $\eqref{eq:embedding-to-imdp}$, and let $\SubPartitionSize_{\max} = \max_{r}L_r$. For $\alpha:=\beta\SizePartition\SizeTargetSets\SubPartitionSize_{\max}$ and for the controller defined by $\phi:=\InputTransSys^*(x,\pi^*(\PartitionRelation(x),k))$ it holds
    \begin{equation}
        \label{eq:confidence-alpha} \min_{\TransitionProbability\in\IntervalTransitionProbability}\ProbabilityMeasure^{\pi^*}_\TransitionProbability\{\varphi'^K_{s_0}\}\geq \eta \Rightarrow \ProbabilityMeasure^N\{\ProbabilityMeasure^{\phi}\{\varphi^K_{x_0}\geq\eta\}\}\geq 1-\alpha. 
    \end{equation}
\end{theorem}

\section{Experimental Evaluation}
\label{sec:experiments}

% -- we can constrcut an abstraction for constrained inputs, which is a limitations for other methods

% -- ther esulting abstraction has many more behaviours, might be difficult to synthesise a policy despite the nondetermninism

% -- if we decide to solve the nondeterminism by taking the min/max probabilities, we might get zero/one probabilities real quick 

Our codebase is based upon  \cite{BRAJ23a}, which 
for brevity is denoted as ``single-target procedure'' (STP), 
and has been modified in order to include multiple-target transitions (hence denoted as MTP). 
% In the following, we compare the abstractions provided by both methods for two test benchmarks. 
% All experiments ran single-threaded on a computer with 4 2.4GHz cores and 8GB RAM.  
The controller synthesis uses the probabilistic model checker PRISM \cite{kwiatkowska2011prism} to compute optimal iMDP policies. Each interval transition probability is computed using $N=20000$ samples and a confidence $\beta = 10^{-8}$. 

\begin{figure}
    \centering
    \includegraphics[trim={1.4cm 0cm 3cm, 0cm}, clip=true, width=0.8\linewidth]{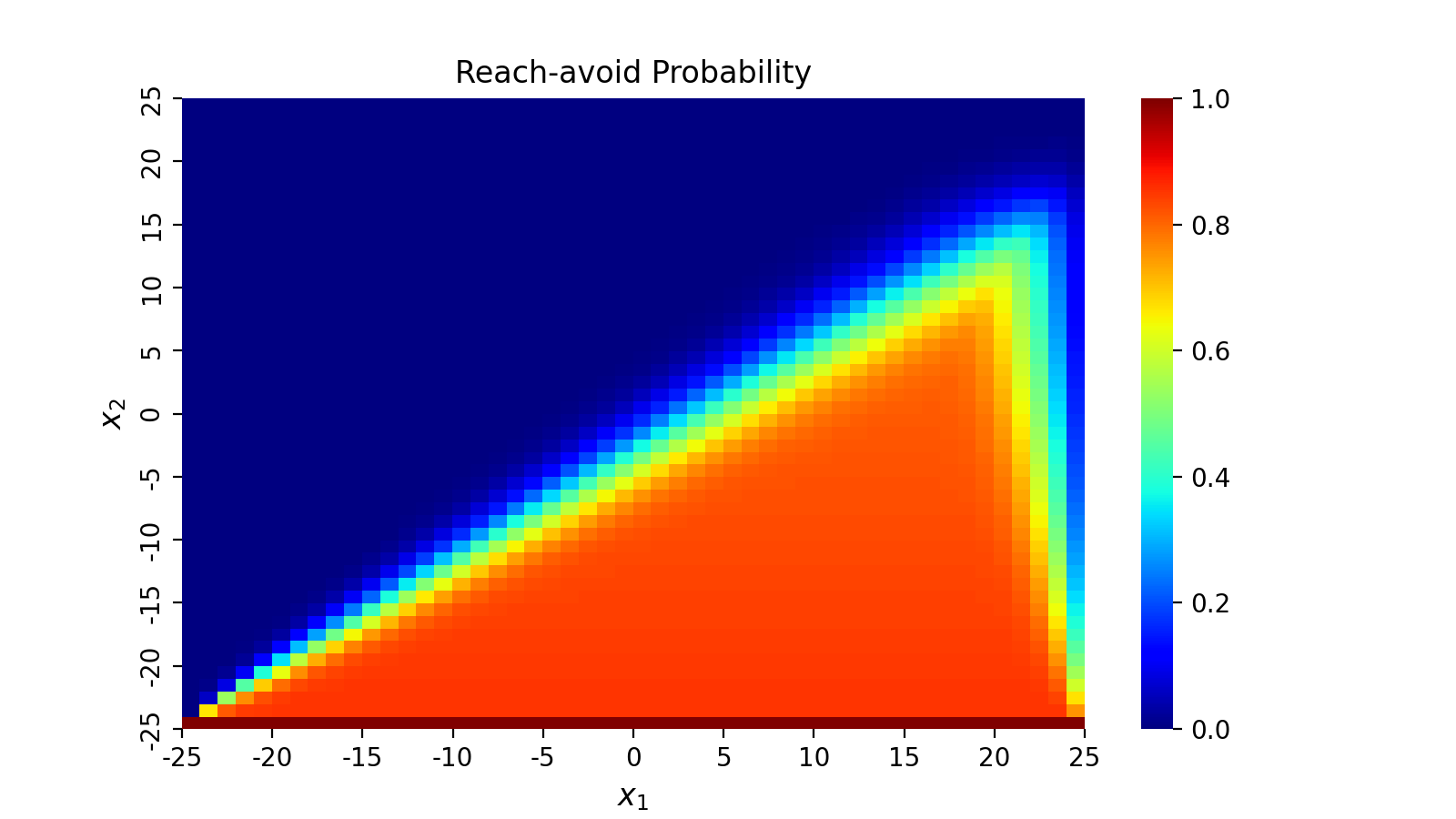}
    \caption{Lower bound on the probability of reaching the goal set (represented by the lowest row of states) for  Example~\ref{exa:motivation}. }
    \label{fig:example-prob}
\end{figure}

\textbf{Example 1 (Cont'd). } 
\label{subsec:example-1-experiments}
Let us consider the dynamical model \eqref{eq:example-dynamics}, over the domain $\Domain = [-25, 25]^2$, partitioned into 2500 square regions, where the goal set is the region $\Domain_{G} = [-25, 25] \times [-25, -24]$. The control input lies in the set $\mathcal{U} = [0, 1] \times [0.5, 1.5]$, and the noise follows a Gaussian distribution $W_k \sim \mathcal{N}(0, 0.15 \cdot I)$. Our goal is the computation of a control policy making the dynamics reach the goal in 50 time steps at most.
As outlined in Example~\ref{exa:motivation}, 
% by considering the transitions occurring from a single region to another single region, 
the STP  returns 
% an empty iMDP, i.e. 
an iMDP with no actions enabled. 
We thus select target sets composed of two adjacent partitions, so that the composite $\Pre$ set covers an entire partition region. 
Our procedure creates an iMDP equipped with 42391 transitions, and computes a policy whose lower bounds on the satisfaction probability is shown in Fig.~\ref{fig:example-prob}, with a confidence (see \eqref{eq:confidence-alpha}) $\alpha \simeq 0.12$. 
% for all states in the domain. 

\textbf{Double Integrator. }
\label{subsec:double-integrator}
Let us consider the stochastic model 
\begin{equation}
    \RVStateTransSys_{k+1} = 
    \begin{bmatrix}
        1 & 1
        \\
        0 & 1
    \end{bmatrix}
    \RVStateTransSys_{k}
    + 
    \begin{bmatrix}
        0.5 & 0 
        \\ 
        0 & 1
    \end{bmatrix}
    \InputTransSys_k 
    + W_k,
\end{equation}
where $W_k \sim \mathcal{N}(0, 0.15 \cdot I)$. The reach-avoid task is to reach set $[-2, 2]^2$ within 5 time steps, while avoiding states $\RVStateTransSys \notin [-11, 11]^2$. The control input is limited by the set $[-2, 4] \times [-3, 3]$. 
We partition the domain $\Domain = [-11, 11]^2$ into square partitions, in four different configurations: 11x11, 15x15, 18x18, 20x20 regions.
%, 33x33 regions. 
We consider target sets composed of two adjacent regions. The complete results are reported in Table~\ref{tab:result-comparison}, in terms of computational time, number of transitions of the resulting abstractions, and in percentage of states with a positive probability of reaching the goal set, along with the confidence $\alpha$ (see \eqref{eq:confidence-alpha}) for the MTP approach.
The first two partitioning choices return an empty iMDP when considering the STP, as the $\Pre$ of a region is not entirely contained in any single partition, whilst the MTP successfully construct an iMDP. 
Starting from the case of 18x18 partitions and up, the STP returns smaller abstractions than the MTP; this is expected, as the MTP implicitly considers significantly more target partitions -- this behaviour is reflected in the higher computational times needed to construct the abstract models. 
%
%
% It is interesting to compare
We notice that 
the MTP has a higher percentage of states with a positive probability of reaching the goal set, as it is equipped with more transitions compared to the STP. 
% the portion of states with a positive probability to reach the goal set: 
% as this MTP is equipped with a larger transition set, a higher percentage of states have nonzero probability of satisfying the property under consideration. 
Further, as the partitions' number increases, and thus the partitions' size narrows, the gap between the STP and the MTP shrinks: the abstraction becomes finer for both procedures, providing a more precise depiction of the concrete underlying system.

% \begin{figure}
%     \centering
%     \includegraphics[width=0.8\linewidth, height=4cm]{Figures/backprop_simplerobot.png}
%     \caption{Backward reachable set of the double integrator example. The red and blue centers project the light red and light blue area, respectively. One partition (thick black lines) is contained in the union of the two backward reachable sets.}
%     \label{fig:backreach-d-integrator}
% \end{figure}

% \begin{figure}
%     \centering
%     \includegraphics[width=0.85\linewidth]{Figures/double_integrator_22x22_default_heatmap.png}
%     \caption{Lower bound on the probability of reaching the goal set (located at the center of the figure) for the double integrator dynamics. }
%     \label{fig:r-av-probability-all}
% \end{figure}

\begin{table}[]
    \centering
    \begin{tabular}{c|cc|cc|cc|c}
    Partition & \multicolumn{2}{c}{Transitions} & \multicolumn{2}{c}{Time  [s]} & \multicolumn{2}{c}{Reach  [\%]} & $\alpha$
     \\  
     & 
    STP &  MTP 
    &
    STP &  MTP 
    &
    STP &  MTP 
    &  MTP
    \\ \hline
    $11^2$ & -- & 1907 & -- & 7.9 & -- & 61.9 & 0.0006
    \\
    $15^2$ & -- & 5592 & -- & 11.2 & -- & 62.2 & 0.002
    \\
    $18^2$ & 6423 & 26511 & 8.4 & 28.6 & 54.3 & 66.0 & 0.01
    \\
    $20^2$ & 13155 & 52580 & 9.8 & 35.1 & 62.5 & 68.5 & 0.03
     \\
     $22^2$ & 22422 & 90896 & 11.5 & 43.7 & 61.1 & 66.9 & 0.05
    %\\
    %$33^2$ & 369121 & 1143719 & 25.9 & 115.4 & 70.3 & 72.4 
    \end{tabular}
    \caption{Comparison between the STP and MTP abstractions, in terms of number of transitions, computational time, and percentage of states that have a positive probability to reach the goal set, and confidence $\alpha$ for the MTP. }
    \label{tab:result-comparison}
%     \licio
% {
%  Nice! Definitely an improvement with respect to what we were doing before, at the expense of generating more complex transition systems. No free lunch. :)
% }
\end{table}

%%%%%%%%%%%%%%%%%%%%%%%%%%%%%%%%%%%%%%%%

\section{Discussion and Conclusions}
\label{sec:conclusion}

% In this letter, 
We have developed a novel abstraction procedure for discrete-time stochastic systems, exploiting nondeterministic transitions to generate finite-state abstract models.  
% generalising the scope of earlier results. 
By allowing each transition to reach a \emph{set} of partitions, rather than a single region, we show that we can build an abstraction even when the backward reach set does not entirely cover any single partition region, thus generalising the scope of earlier results. 
This event can occur for systems whose input set is constrained (e.g. with significant saturation): thus, our  approach is suited for control designs with performance costs. 
Our experimental evidence shows that this flexibility comes at the cost of generating larger (in terms of transitions) models than the existing single-target approach, and hence introducing more behaviours in the abstraction. 
The selection of target sets and the embedding of an RMDP into an iMDP affect the performance of our method: as such, a deeper study of tailored algorithms for RMDPs which exploit the structure of the uncertain transition function obtained by this scheme are matter of future efforts.

% conference papers do not normally have an appendix

% use section* for acknowledgment

% \section*{Acknowledgment}
% The authors would like to thank...

% trigger a \newpage just before the given reference
% number - used to balance the columns on the last page
% adjust value as needed - may need to be readjusted if
% the document is modified later
%\IEEEtriggeratref{8}
% The "triggered" command can be changed if desired:
%\IEEEtriggercmd{\enlargethispage{-5in}}

% references section

% can use a bibliography generated by BibTeX as a .bbl file
% BibTeX documentation can be easily obtained at:
% http://mirror.ctan.org/biblio/bibtex/contrib/doc/
% The IEEEtran BibTeX style support page is at:
% http://www.michaelshell.org/tex/ieeetran/bibtex/
%\bibliographystyle{IEEEtran}
% argument is your BibTeX string definitions and bibliography database(s)
%\bibliography{IEEEabrv,../bib/paper}
%
% <OR> manually copy in the resultant .bbl file
% set second argument of \begin to the number of references
% (used to reserve space for the reference number labels box)
\bibliographystyle{abbrv}
\bibliography{library}

\end{document}